\documentstyle[12pt,epsfig]{article}
\newcommand{\beq}{\begin{equation}}
\newcommand{\eeq}{\end{equation}}
\newcommand{\bea}{\begin{eqnarray}}
\newcommand{\eea}{\end{eqnarray}}

\newcommand{\ra}{\rightarrow}

\newcommand{\half}{\mbox{$\frac{1}{2}$}}

\newcommand{\gsim}{\lower.7ex\hbox{$
\;\stackrel{\textstyle>}{\sim}\;$}}
\newcommand{\lsim}{\lower.7ex\hbox{$
\;\stackrel{\textstyle<}{\sim}\;$}}

\def\lsim{\mathrel{\rlap{\lower3pt\hbox{\hskip0pt$\sim$}}
    \raise1pt\hbox{$<$}}}         
\def\gsim{\mathrel{\rlap{\lower4pt\hbox{\hskip1pt$\sim$}}
    \raise1pt\hbox{$>$}}}         

\newcommand{\bibit}[1]{\bibitem{#1}}

\newcommand{\aver}[1]{\langle #1\rangle}

\newcommand{\La}{\overline{\Lambda}}
\newcommand{\Lam}{\Lambda_{\rm QCD}}

\newcommand{\as}{\alpha_s}
\newcommand{\GeV}{\,\mbox{GeV}}
\newcommand{\MeV}{\,\mbox{MeV}}
\newcommand{\matel}[3]{\langle #1|#2|#3\rangle}
\newcommand{\state}[1]{|#1\rangle}

\begin{document}

\begin{titlepage}
\renewcommand{\thefootnote}{\fnsymbol{footnote}}

\begin{flushright}
UND-HEP-00-BIG\hspace*{.04em}13\\
hep-ph/0012336
\end{flushright}
\vspace*{1.3cm}

\begin{center} \Large
{\bf A Few Aspects of Heavy Quark Expansion}
\end{center}
\vspace*{1.3cm}
\begin{center} {\Large
Nikolai Uraltsev} \\
\vspace{1.2cm}
{\normalsize
{\it Dept.\ of Physics, Univ.\ of Notre Dame du Lac, Notre Dame, IN 46556,
U.S.A.$^*$}\\
{\small \rm and} \\
{\it INFN, Sezione di Milano, Milan, Italy$^{\:*}$
}\\
}
\normalsize \vspace*{5mm}

\vspace*{1.2cm}

{\small 
{\sf To appear in the Proceedings of the UK Phenomenology Workshop} on\\ 
{\it Heavy Flavour and CP Violation}\\ {\rm Durham, UK, 17--22 September
2000}
}
\vspace*{1.2cm}

{\large{\bf Abstract}}\\
\end{center}
\vspace*{.2cm}
Two topics in heavy quark expansion are discussed. The heavy quark potential 
in perturbation theory is reviewed in
connection to the problem of the heavy quark mass. The nontrivial reason 
behind the
failure of the ``potential subtracted'' mass in higher orders is
elucidated. The heavy quark sum rules are the second subject. The
physics behind the new exact sum rules is described and a simple quantum
mechanical derivation is given. The question of saturation of sum rules
is discussed.  A comment on the nonstandard possibility which would affect 
analysis of ${\rm BR}_{\,\rm sl}(B)$ vs.\ $n_c$ is made.

\vfill

~\hspace*{-7mm}\hrulefill \hspace*{3cm} \\
\footnotesize{\noindent $^*$On leave of absence from 
St.\,Petersburg Nuclear Physics Institute, Gatchina, St.\,Petersburg 
188300, Russia}

\noindent
\end{titlepage}

\newpage

\section{Introduction}

Heavy quark symmetry and the heavy quark expansion have played an
important role in understanding weak decays of heavy flavors. Recent
years witnessed significant success in quantifying strong
nonperturbative dynamics in a number of practically important problems
via application of Wilson Operator Product Expansion (OPE).
The advances were in theoretical understanding of strong
dynamics of heavy quarks, as well as in practical applications to
phenomenology of beauty decays. The discussion of both aspects can be
found in the recent review \cite{ioffe}. It also included updates on such
questions of general interest as the current theoretical uncertainties
in extracting the KM parameters $|V_{cb}|$ and $|V_{ub}|$, our knowledge
of the heavy quark masses and other basic parameters of the heavy quark
theory, aspects of local quark-hadron duality relevant to the
quest for $|V_{cb}|$ and $|V_{ub}|$. The interested reader can find
there references to the original publications.

In this contribution I address a few selected topics which have not been
spelled in much detail in the literature. Even though some are of more
theoretical nature, I try to place emphasis on the general physical
features understandable without presenting much of the technicalities. I
discuss in some detail the notion of the heavy quark potential in QCD
and its limitations, in particular as viewed through its relation to the 
heavy quark mass. The origin of the subtleties is explained which led to 
a failure
in defining the low-scale running heavy quark mass in higher orders
using the usual interquark potential. 

Another topic is the heavy quark sum rules, where the significant
part is dedicated to new exact spin sum rules; applications to
phenomenology are discussed. I conclude with the brief comment on charm
counting in $B$ decays.

\section{Heavy quark potential and the heavy quark mass}

Strictly speaking, the heavy quark potential appears in a somewhat different
situation compared to usual $B$ decays, namely where heavy quark and 
antiquark (with small relative velocities) 
are present simultaneously. It attracted recently renewed attention 
in connection with the pair production of $t\bar{t}$ and $b\bar{b}$
near threshold, and through its connection with the problem of heavy 
quark mass. 
The review of applications for the $Q\bar{Q}$ system can be found, for 
example, in Ref.~\cite{nora}. Here we give a more pedagogical discussion of
the underlying problems.

A closer look at the notion of the heavy quark potential in QCD reveals 
certain subtleties. It turns out that a literal analogue of potential 
interaction does not exist in QCD. 

The original notion of the potential refers to the interaction of
infinitely heavy (static), or completely nonrelativistic heavy
particles, which is {\em instantaneous}. The most familiar example is the
electromagnetic interaction of heavy charges. The Hamiltonian
of such a potential system is given by
\beq
{\cal H} = \sum_i \frac{(i\vec\partial\,)^2}{2m_i} +
V\left(\vec{r}_i\!-\!\vec{r}_j\right)\;,
\label{pot5}
\eeq
where $m_i$ are masses of particles and the potential $V$ is a function
of their instant  coordinates. Taking the limit $m_i\!\to\!
\infty$ (at fixed $\vec r_i$, which corresponds to semiclassically high
excitation numbers of the quantum system in Eq.\,(\ref{pot5}))
eliminates quantum uncertainties in the coordinates and allows to
measure the potential directly as the position-dependent energy of the 
infinitely slowly
moving collection of particles. This is well-known for QED where the
potential of the charges $q_1$ and $q_2$ (in units of 
the electron charge)
is given by
\beq
V^{\rm QED}(R) = \alpha_{{\rm em}}(0) \,\frac{q_1 q_2}{R}
\;.
\label{pot7}
\eeq
This expression is exact in the absence of light charged particles; 
known quantum corrections appear only if other matter
fields are not much heavier than the scale $1/R$.

The definition of the similar quantity in QCD turns out more tricky, since 
color sources are gauge-dependent. 
The color of the
individual heavy quark remains fully quantum in nature and changes
through  interaction with gluons. In contrast to usual coordinates,
the limit $m_Q\!\to \!\infty$ does not make color variables describing the
state of the heavy quark semiclassical.

To avoid this problem, the heavy quark potential is usually defined via the
vacuum expectation value of stretched  Wilson loops,
\beq
V(R) = -\lim_{T\to\infty}\, \frac{1}{T} \ln{\langle{\rm Tr}
\:{\cal P}\, {\rm exp}
\left(
i\oint_{C(R,T)} A_\mu {\rm d}x_\mu
\right)}
\rangle \;,
\label{pot11}
\eeq
where the rectangular contour $C(R,T)$ spans distances $R$ and $T$ in the
space and time directions, respectively. (It is usually assumed to be in
Euclidean space as in Eq.\,(\ref{pot11}).)  The gauge field is taken in the 
color representation of the heavy quark (fundamental representation for 
actual quarks). 
This definition is
intuitively clear, since such Wilson lines describe propagation of 
infinitely heavy quarks. Moreover, such Wilson loops are readily
computed in the U(1) gauge theory (free QED) and, of course,
reproduce Eq.\,(\ref{pot7}) (with $q_1\!=\!-q_2$).
In QCD the perturbative expansion of $V(R)$ has been computed through
order $\alpha_s^3$.\cite{schroder} Usually the potential in the
momentum representation is considered,
\beq
V(\vec{q}\,) = \int {\rm d}^{3\!} \vec R \; V(\vec R) \; {\rm
e}\,^{-i\vec{q}\vec{R}}
= -\frac{4}{3} \frac{4\pi\alpha_s}{\vec{q}^{\:2}}
\left\{\!1\!+ 
\!\left(\!\frac{31}{3}\!-\!\frac{10}{9}n_{\!f}\!\right) \frac{\alpha_s}{4\pi} 
\!+ \!c_3
\!\left(\!\frac{\alpha_s}{4\pi}\!\right)^{\!2} \!\!+... \right\}.
\label{pot13}
\eeq

Taking the definition (\ref{pot11}) for $V(R)$, one faces a number of 
questions. What type of processes does $V(R)$ incorporate and what are its 
properties, say, in the perturbative expansion? Can it be used, say, in the 
Schr\"{o}dinger equation similar to usual quantum mechanics or QED? It turns 
out that both questions are nontrivial and interrelated.

There are important peculiarities in the thus defined static interaction in the
$Q\bar{Q}$ system; they were first analyzed by Appelquist {\it et al.}\ 
already in the late 1970's \cite{appel}. Due to 
the gluon self-interaction,
the potential  contains, in higher orders, the ${\rm H}$-type diagrams 
shown in
Figs.~1. Here the dashed line denotes the Coulomb quanta (they mediate
instantaneous interaction in the physical Coulomb gauge), while the wavy
lines are used for the transverse gluons. These diagrams  
propagate in time the transverse gluons. This means that at this level 
the problem ceases to be a two-body one, and includes more full-fledged
dynamical degrees of freedom. 

\begin{figure}[hhh] 
 \begin{center} \vspace*{-2mm}
 \mbox{\epsfig{file=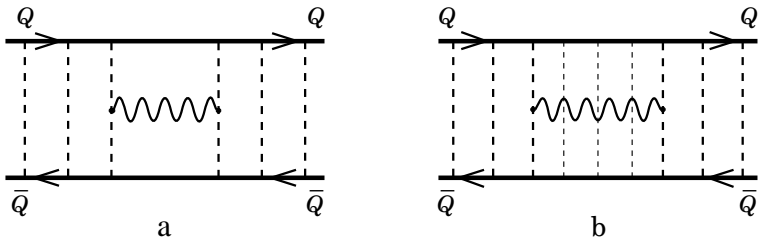,width=10.0cm}} \vspace*{-6mm}
  \end{center}
\caption{
Examples of diagrams for heavy quark potential in QCD. Dashed lines are
instantaneous Coulomb exchanges, transverse gluons propagating in time 
are shown by wavy lines.  
a)~The convergent ${\rm H}$-diagrams with the transverse gluon as a rung. 
b)~Adding more Coulomb exchanges inside the ``${\rm H}$'' 
leads to infrared divergence in perturbation theory.
}\vspace*{-2mm}
\end{figure}

Moreover, while the first ${\rm H}$-diagram Fig.~1a appearing
in order $\alpha_s^3$ safely converges at the rung gluon momenta $\sim \!1/R$,
including additional Coulomb exchanges, as in Fig.~1b leads to an 
infrared divergence. With one exchange it is logarithmic; the 
formal degree of the infrared divergence increases
with adding extra Coulomb quanta between the emission and absorption of
the transverse gluon. Physics behind this infrared behavior was
discussed in Ref.~\cite{appel}:  emission of the 
soft transverse gluon
changes the overall color of the $Q\bar{Q}$ pair and, therefore, modifies
the interaction energy between them. The energy shift
associated with the exchange of the transverse gluon depends
nonanalytically on the energy denominator (in the language of
noncovariant time-ordered perturbation theory), since the gluon can be
arbitrarily soft. Formally expanding the exact result in $\alpha_s$
includes expanding in this change of the Coulomb energy proportional to
$\alpha_s/R$. Therefore, one
obtains increasing infrared singularities. These arguments suggested
that resummation of the Coulomb exchanges in Fig.~1b would render these
diagrams finite. However, the effective infrared cut-off is of the order
of $\alpha_s/R$ and, thus, the potential is not 
 {\em
perturbatively} infrared finite 
 containing terms $\sim 1/R\cdot \ln{\alpha_s}$ starting at the
order ${\cal O}\left(\alpha_s^4\right)$.

This purely perturbative analysis shows that there is no direct analogue
of the potential between heavy quarks in QCD. The $Q\bar{Q}$ system
incorporating all gluon interactions is not a two-body system but
includes actual propagation of gluons with energy small compared to
$1/R$. The interaction, then, cannot be universally described by an
instantaneous potential and is intrinsically nonlocal in time. 
Thus, the prescription (\ref{pot11}) based on Wilson
loops in QCD does not yield the heavy quark potential in its
conventional understanding. Theoretical treatment of the 
$Q\bar{Q}$ system has to account for these peculiarities of QCD. Examples of 
the technique used here can be found in Ref.~\cite{nora} and other papers 
referred to there.

In spite of all subtleties mentioned above, the heavy quark potential $V$ 
defined via Wilson loops, Eq.\,(\ref{pot11}), is an observable quantity (at 
least in the perturbative regime $R \ll 1/\Lam$), up to an overall additive 
constant appearing in the renormalization of the straight Wilson line. In 
particular, it is gauge-invariant in perturbation theory. It is therefore 
tempting to use it to quantify various strong interaction effects, 
including low-energy contribution to the mass of the heavy quark.

\begin{figure}[hhh]
 \begin{center} \vspace*{-1mm}
 \mbox{\epsfig{file=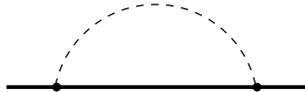,width=4cm}}
  \end{center}\vspace*{-6mm}
 \caption{Self-energy diagram yielding the classical Coulomb shift in the 
mass of a nonrelativistic particle.
}\vspace*{-2mm}
\end{figure}

It is well known that in classical electrodynamics the self-energy of a  
charged particle is given by the potential at the origin, 
$\frac{1}{2} e^2 V(0)$. 
The same holds in quantum electrodynamics in respect to the 
effect of quanta with 
momenta much smaller than the source mass (or inverse radius, for the 
composite particle), as is illustrated by the diagram Fig.~2 yielding 
\beq
\delta m \simeq
e^2 \int \frac{d^3\vec k}{4\pi ^2} \frac{\alpha}{\vec k^2}
= \frac{1}{2}\, e^2\, V_{\rm em}(0)
\; .
\label{3.2a}
\eeq
Here $e$ is charge and $V_{\rm em}$ is the potential between 
the same-sign charges, hence it is 
positive. In QCD the perturbative
diagrams for both $m_Q$ and $V(\vec{q}\,)$ are more complicated. Yet, the
similar relation holds in order ${\cal O}(\alpha_s^2)$ as well. 
This is most simply seen in the Coulomb gauge 
where all the effect at this order
reduces to dressing the propagator of the Coulomb quanta, i.e. using the 
running $\alpha_s(\vec{q}^{\,2})$. (An
alternative discussion can be found in Ref.~\cite{beneke}.)
There is a general argument \cite{ioffe} showing 
that such a relation holds to all orders in
perturbation theory: the infrared contribution to $V(0)$,  
\beq
V_{\rm IR}(0) = \int \frac{{\rm d}^3 \vec q}{(2\pi)^3} 
\: V_{\rm IR}(\vec{q}\,) \;.
\label{pot21}
\eeq equals to minus twice the same contribution to the mass of a
static color source.

The general structure of the perturbative diagrams for heavy quark
mass and $V(0)$ indeed is similar. To establish the correspondence,
one can cut the $\bar{Q}$ line in the diagram for the potential at
some place and turn the new external legs around. Integration over
momentum $\vec q$ corresponds to closing the original external $Q$ and
$\bar Q$ legs, as exemplified by Fig.~3.  This correspondence often
works out in a nontrivial way. For example, in the potential we should
discard the ``reducible'' diagrams which can be cut across only $Q$
and $\bar Q$ lines, Fig.~4.a. The corresponding contributions in
$\delta m_Q$ would include rainbow diagrams like in Fig.~4b. The
latter, however, vanish in the Coulomb gauge, which is self-manifest
in the coordinate representation. The Coulomb propagator is
instantaneous, $\delta(t_1\!-\!t_2)/|\vec{x}\!-\!\vec{y}\,|$ while
the heavy quark propagator is retarded and includes
$\theta(t_1\!-\!t_2)$. In diagrams like Fig.~4b we have $\tau_2\!>\!\tau_1$
and therefore they vanish. The exception is one-loop diagram which
does not vanish since in the integral over $\omega$ the large
semicircle at $\omega \to \infty$ yields the finite contribution; this
is equivalent to the prescription $\theta(0)\!=\!\frac{1}{2}$.

\begin{figure}[hhh]
 \begin{center} \vspace*{-2mm}
 \mbox{\epsfig{file=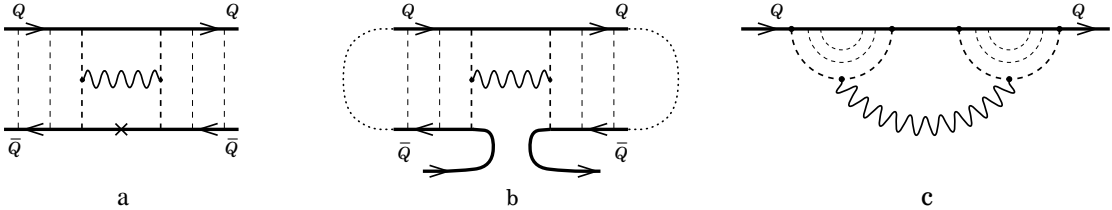,width=14.7cm}} \vspace*{-6mm}
  \end{center}
 \caption{
Correspondence between the diagrams for potential $V(0)$ and for the heavy 
quark mass $m_Q$. The dotted lines denote closing the external quark lines and 
integrating over $\vec{q}$.
}\vspace*{-2mm}
\end{figure}

The general argument goes as follows. Let us imagine we were able to
introduce in some way an ensemble of gauge field configurations where
the modes with momenta much larger than a certain scale $\mu$ are
practically absent. This field-theoretic system would not need
regularization, and everything can be expressed in terms of the bare
parameters, including the bare quark mass $m_Q^{(0)}$.  At $R\!\to\!
0$ the Wilson loop will approach its free value $N_c$ thus yielding
$V(0)\!=\!0$. This is clear on physical grounds: $Q\bar{Q}$ form a
dipole with the infinitesimal dipole moment, and its interaction with
any soft gluon field vanishes as $R$ goes to $0$. It is important at
this point that our gauge ensemble explicitly includes only soft
modes. Otherwise, as in full QCD, the modes with $|\vec{k}\,| \sim
1/R$ generate growing attractive potential at arbitrary small $R$.

\begin{figure}[hhh]
 \begin{center} \vspace*{-0mm}
 \mbox{\epsfig{file=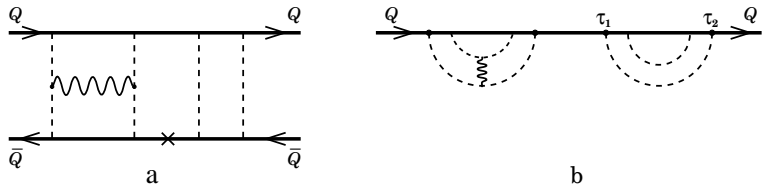,width=10cm}} \vspace*{-6mm}
  \end{center}
 \caption{
Reducible $\bar{Q}Q$-only diagrams are not included in $V(R)$ being  
iterations of the potential interaction. Similar diagrams in $m_Q$ vanish due 
to instantaneous nature of the Coulomb interaction.
}\vspace*{-2mm}
\end{figure}

Next, we note that $V(R)$ is traditionally determined up to a
constant.  In perturbation theory (in four dimensions) the potential
is actually defined as \beq U(R) = V(R)-V(\infty) \,.
\label{pot23}
\eeq We assign $U$ to this ``standard'' potential to distinguish it
from $V(R)$ which has a precise meaning in a finite theory. While $U(R)$
by definition vanishes at $R\to \infty$, $V(\infty)$ does not and
reflects nontrivial interaction with the gluon field. In the momentum
representation $V(\vec{q}\,)$ explicitly contains the term
$V(\infty)\,\delta^3(\vec{q}\,)$, which is discarded in standard
perturbative computations. Since at $R\!\to \!\infty$ the $\bar{Q}$
and $Q$ lines are well separated, their interaction must vanish as
$1/R$, (at least, in perturbation theory) and the value of the Wilson
loop is given by the mass renormalization of each static source, \beq
V(\infty) = 2\left(m_Q^{{\rm ren}} - m_Q^{(0)}\right) = 2\delta m_Q
\;.
\label{pot25}
\eeq Therefore, we have for the ``ordinary'' potential \beq U(0) =
\int\frac{{\rm d}^3 \vec q}{(2\pi)^3} \: V_{\rm reg}(\vec{q}\,) =
-2\delta m_Q \;.
\label{pot27}
\eeq Here $V_{\rm reg}$ is the regular part of $V(\vec{q}\,)$
(computed in a usual way) not containing self-energy diagrams yielding
$\delta^3(\vec{q}\,)$.  Thus, we have the stated relation between
$V_{\rm IR}(0)$ and $\delta_{\rm IR}m_Q$.

Relying on relation (\ref{pot27}), one can try to perturbatively
define a certain running heavy quark mass\footnote{This
idea was discussed by myself in 1996 and later, independently,
advocated by M.~Beneke \cite{beneke}.} which is free from the
leading renormalon uncertainty $\sim \Lambda_{\rm QCD}$ \cite{pole}.  
One defines
\beq 
m_Q^{\rm PS}(\mu) = m_Q^{\rm pole} +
\half \int_{|\vec{q}\,|< \mu} \mbox{$\frac{{\rm d}^3 \vec
q\;}{(2\pi)^3}$} \: V_{\rm reg}(\vec{q}\,)
=\; m_Q^{\rm pole} + \mbox{$\frac{1}{\pi}$} \int_0^{\infty} \!\!{\rm
d}R\; V(R) \!\left[ \mbox{$\frac{\sin{\mu R}}{R}$} \!-\! \mu\cos{\mu
R} \right] ,
\label{pot31}
\eeq 
where $V(\vec{q}\,)$ is computed to a certain order in
perturbation theory with no explicit cut-off, and the pole mass is
taken to the same order in $\alpha_s$. The mass $m_Q^{\rm PS}(\mu)$ is
known as the ``potential-subtracted'' mass \cite{beneke}.  As is clear
from the preceding derivation, this mass would have a meaning of the
rest-frame energy of the heavy quark, but not the mass determining the
kinetic energy, which is generally different once a cutoff is
introduced (i.e., $m_0$ rather than  $m_2$ in the notations of
Refs.~\cite{varenna,ioffe}).

The {\em ansatz} (\ref{pot31}) is supposed to remove 
from the pole mass all infrared contributions originating from the scale 
much below the cutoff $\mu$. As we saw, this happens to one and two loops. 
Unfortunately, this is not true starting order $\alpha_s^3$.

The problem becomes self-manifest at order $\alpha_s^4$ where 
infrared-singular contributions of Fig.~1b emerge in $V(\vec{q}\,)$ 
which behave 
like $\frac{\alpha_s^4}{\vec{q}^{\,2}}
\ln{\frac{\vec{q}^{\,2}}{\epsilon}}$, with $\epsilon$ being an infrared cutoff
in the ``rung'' gluon momentum. On the other hand, the corresponding 
contributions 
are absent from $m_Q$ since the Coulomb exchanges are instantaneous.

This apparent contradiction does not mean that the relation
(\ref{pot27}) is violated. As explained in Ref.~\cite{ioffe}, the
subtlety resides in the necessity to introduce the cut-off on the
gluon momenta to compute $V_{{\rm IR}}$. Here we consider this problem
from the different perspective.

Let us take a look at diagram Fig.~3a. It does contribute to $V(0)$
computed via the simple-minded prescription in Eq.\,(\ref{pot31}). At
the same time the diagram Fig.~3c vanishes simply since the one-loop
$\bar Q Q g$ vertex is zero for transverse gluon. The {\em ansatz}
(\ref{pot31}) treats the diagrams in Figs.~3a and 3c differently: the
integrals run over all momenta of the Coulomb quanta in the latter,
but there is a nontrivial cut at momenta of order $\mu$ for potential in
Fig.~3a. Our general proof actually states that the amplitudes of
emitting the transverse gluon must vanish once the incoming $Q$ and
$\bar Q$ lines are contracted performing the integration over $\vec
q$. This is the cancellation of the diagrams in Fig.~5, which are
subdiagrams to the whole potential $V(0)$. Clearly, the cancellation
holds only if the cuts in all diagrams are the same, which is 
not the case in the {\em ansatz} (\ref{pot31}).

\vspace*{-2mm}
\begin{figure}[hhh]
 \begin{center}
 \mbox{\epsfig{file=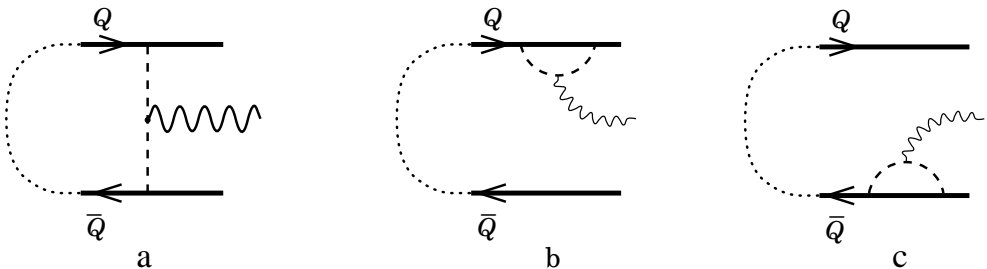,width=13.cm}} \vspace*{-6mm}
  \end{center}
 \caption{
Diagrams for interaction of the transverse gluon which cancel upon 
integration over $\vec{q}$ in the absence of hard Coulomb quanta. 
The cancellation, however does not occur for potential in 
Eq.~(10).
}\vspace*{-2mm}
\end{figure}

The above consideration reveals the problem of the definition of the
``potential subtracted'' mass at the technical level. In fact, there
is a deeper reason behind its failure, 
which is instructive. The purpose of the additional term in Eq.\,(\ref{pot31}) 
is to remove the infrared
contributions from the pole mass. If the gluons in the diagrams were
all soft with momenta not exceeding certain scale $\epsilon$, the
induced potential in the momentum representation would vanish
completely above some scale $\mu \gg \epsilon$. Then
$\int_{|\vec{q}\,|<\mu} {\rm d}^3 \vec{q}\; V_\epsilon(\vec q\,)$ would
comprise this contribution completely and correctly, regardless of
any technical details. Why then does {\em ansatz} (\ref{pot31}) fail
in practice?

The reason lies in presence of both soft ($\vec q \!\lsim\! \epsilon$) and
hard ($\vec q \!\sim \!\mu$) gluons in the potential, and in lack of necessary
``factorization'' of these scales in $V(\vec{q}\,)$. Let us single out
the soft transverse gluon with momentum $\vec{k} \!\sim\!
\epsilon$. If it affected the full potential only at $|\vec{q}\,|
\lsim \epsilon$, everything would work fine.  However, as exemplified
by the ${\rm H}$-diagrams in Fig.~1, this is {\it not} the case. Presence of
hard gluons propagates the contribution of the soft gluon to all
Fourier components of $V(\vec{q}\,)$. The impact of hard gluons on processes with soft transverse gluons is not limited to only renormalizing their bare 
interactions, but also introduces new nontrivial `dipole' matrix
elements between the heavy quark states, which, for example, do not
vanish when the momentum $\vec k$ goes to zero.

The definition of the ``kinetic'' running heavy quark mass based on
the small velocity sum rules is protected against such problems by the
operator product expansion (OPE). The effect of any soft gluon on the
moments of the SV structure functions is given by the corresponding
heavy quark operators whether or not hard gluons are
present. Likewise, the OPE bars soft physics from being transferred to
large energy -- the corresponding effects are suppressed by at least
third power of energy \cite{dipole} and are described by the OPE.

The lesson one can draw from the analysis of the heavy quark
potential/mass problem is that extreme care must be exercised in
treating the infrared safe observables for which the OPE does not
apply.

\section{Heavy quark sum rules}

An important class of constraints on the hadronic parameters determining 
the properties of heavy flavor hadrons follow from the heavy quark 
sum rules, among which the small velocity (SV) sum rules in the static 
limit $m_Q \!\to\! \infty$ play a special role. These sum rules are
\bea
\varrho^2\!-\!\frac{1}{4}\!\! &=& 
 2\sum_m\;|\tau_{3/2}^{(m)}|^2 \,\;\;\;+\;\;\;
\sum_n\;|\tau_{1/2}^{(n)}|^2 \, ,
\label{bj}
\\
\,\;\;\;\frac{1}{2}\:\:  &=& 
 2\sum_m\;|\tau_{3/2}^{(m)}|^2 \,\;\;\;-\:
 2\sum_n\;|\tau_{1/2}^{(n)}|^2 \, ,
\label{ur}
\\
\,\;\;\;\frac{\La}{2}\,\, &=& 
2\sum_m \epsilon_m|\tau_{3/2}^{(m)}|^2 \;+\;\;\;
\sum_n \epsilon_n|\tau_{1/2}^{(n)}|^2\, ,
\label{vol}
\\
\;\;\;\;\overline \Sigma\:\, &=& 
2\sum_m \epsilon_m|\tau_{3/2}^{(m)}|^2 \;-\;
2\sum_n \epsilon_n|\tau_{1/2}^{(n)}|^2\, ,
\label{ur2}
\\
\;\;\;\frac{\mu_\pi^2}{3} &=&
2\sum_m \epsilon_m^2|\tau_{3/2}^{(m)}|^2 \;+\;\;\;
\sum_n \epsilon_n^2|\tau_{1/2}^{(n)}|^2\, ,
\label{srmupi}
\\
\;\;\;\frac{\mu_G^2}{3} &=&
2 \sum_m \epsilon_m^2|\tau_{3/2}^{(m)}|^2 \;-\;
2 \sum_n \epsilon_n^2|\tau_{1/2}^{(n)}|^2 \, ,
\label{pig}
\\
\;\;\;\frac{\rho_D^3}{3} &=& 
2\sum_m \epsilon_m^3|\tau_{3/2}^{(m)}|^2 \;+\;\;\;
\sum_n \epsilon_n^3|\tau_{1/2}^{(n)}|^2\, ,
\label{fourth}
\\
-\frac{\rho_{LS}^3}{3}\! &=&
2 \sum_m \epsilon_m^3|\tau_{3/2}^{(m)}|^2 \;-\;
2 \sum_n\epsilon_n^3|\tau_{1/2}^{(n)}|^2 \, ,
\label{fourthls}
\eea
a sequence which, in principle, can be continued further. Here
$\epsilon_k$ is the excitation energy of the $k$-th intermediate state
(``$P$-wave states" in the quark-model language),
$$
\epsilon_k=M_{H_Q^{(k)}}-M_{P_Q}\;  ,
$$
while the functions $\tau_{3/2}^{(m)}$ and $\tau_{1/2}^{(n)}$ describe
the transition amplitudes of the ground state $B$ meson to these
intermediate states. We follow the notations of 
Ref.~\cite{isgw},
\beq
\frac{1}{2M_{H_Q}}\langle H_Q^{(1/2)} |A_\mu |P_Q\rangle
= - \tau_{1/2} \,(v_1\!-\! v_2)_\mu\, ,
\eeq
and
\beq
\frac{1}{2M_{H_Q}}\langle H_Q^{(3/2)} |A_\mu |P_Q\rangle
= -\frac{1}{\sqrt{2}}\,i\,\tau_{3/2} \,\epsilon_{\mu\alpha\beta\gamma}
\,\varepsilon^{*\alpha} \,v_2^\beta \,v_1^\gamma\, ,
\eeq
where 1/2 and 3/2 mark the quantum numbers of the light cloud in
the intermediate states, $j^\pi = 1/2^+$ and $3/2^+$, respectively,
and $A_\mu$ is
the axial current. Furthermore, the slope parameter $\varrho^2$ of
the Isgur-Wise function is defined as
\beq
\frac{1}{2M_{P_Q}}
\matel{P_Q(\vec v)}{\bar Q \gamma_0 Q}{P_Q}=
1-\varrho^2\,\frac{\vec v^{\,2\!}}{2} +{\cal O}(\vec v^{\,4})\, .
\label{slope}
\eeq

Equation (\ref{bj}) is known as the Bjorken sum rule \cite{BJSR}.
Superconvergent sum rules (\ref{ur}) and (\ref{ur2}) are new 
\cite{ioffe,newsr}. 
Equation (\ref{vol}) was obtained by Voloshin \cite{volopt}. The
expression for kinetic expectation value $\mu_\pi^2$ is the BGSUV 
sum rule \cite{third}. The next
one for chromomagnetic operator was derived in Ref.~\cite{rev}, as well 
as Eq.\,(\ref{fourthls}).
The last two sum rules are obtained along the same lines. 
The sum rule for the Darwin term  $\rho_D^3$ 
was first presented in Ref.~\cite{pirjol}. 
We have introduced the new parameter of the heavy quark theory 
$\overline\Sigma$; it is the small-velocity elastic transition 
matrix element between the states with explicit spin of light degrees of
freedom:
for the ground-state vector mesons such as $B^*$ 
\beq
\frac{1}{\!2M_{\!B^*}\!\!}
\matel{H_Q(\vec{v},\!\varepsilon')}{\bar{Q} iD_{\!j} Q(0)}
{H_Q(0,\!\varepsilon)}  =
- \frac{\La}{2} v_j (\vec\varepsilon\,'^* \vec \varepsilon\,)
-\!
\frac{\overline\Sigma}{2}\! 
\left\{\varepsilon'^*_j(\vec\varepsilon \,\vec v\,)\!-\!
(\vec\varepsilon\,'^* \vec v\,)\varepsilon_j
\!\right\} 
 + 
{\cal O}\!\left(\!\vec v^{\,2}\!\right)\!
.
\label{newsr7}
\eeq
The exact magnitude of the nonperturbative hadronic parameter 
$\overline\Sigma$ is not
known at the moment. Comparing the sum rules
(\ref{ur}), (\ref{ur2}) with the sum rule (\ref{pig}) for the chromomagnetic
expectation value $\mu_G^2$ 
we expect $\overline\Sigma$ to be about $0.25\GeV$.  In the
nonrelativistic system $\overline\Sigma$ is given by the product of
the light mass and the orbital momentum, and would vanish for the
ground-state mesons.

The sum in the r.h.s.\ of Eq.\,(\ref{ur2}) was previously considered in
the recent paper \cite{leya}. It was shown to determine one of the
subleading $B\!\to\! D^*$ formfactors near zero recoil.

The new spin sum rules (\ref{ur}) and (\ref{ur2}) are convergent and are
not renormalized by perturbative effects. This distinguishes them from
all other heavy quark sum rules. The unified derivation of the sum rules
in the field-theoretic OPE is described in detail in the dedicated
papers \cite{optical}, with their quantum mechanical interpretation 
elucidated. A more pedagogical derivation can be found in recent reviews
\cite{rev,ioffe}. The new sum rules were derived in Ref.~\cite{newsr}
applying the OPE to the nonforward SV scattering amplitude off the heavy
quark. Below we will give an alternative quantum-mechanical derivation.
Let us, however,  first discuss physics behind the first sum rule
(\ref{ur}) which is the relation for the total angular momentum of light
cloud in $B$ meson.

At first sight sum rule (\ref{ur}) which is independent of the strong
dynamics looks surprising. In the quark models the
$\frac{1}{2}$- and $\frac{3}{2}$-states are differentiated only by
spin-orbital interaction. The latter naively can be taken arbitrarily
small if the light quark in the meson is nonrelativistic. To
resolve this apparent paradox we note that in the
nonrelativistic case $\tau^2$ are large scaling like inverse square of
the typical velocity of the light
quark, $\tau^2\!\sim\! 1/\vec v_{\rm sp}^{\,2}$. The relativistic
spin-orbital effects must appear at the relative level
$\sim \!\vec v_{\rm sp}^{\,2}$ because spin ceases to commute with
momentum to this accuracy due to Thomas precession. The latter
phenomenon lies behind the sum rule (\ref{ur}). 
This connection will be elucidated below.
The above relativistic corrections lead to the terms of order $1$ in the
first sum rules Eqs.\,(\ref{bj}), (\ref{ur}).

To understand the connection of the sum rule (\ref{ur}) with spin, let us
consider a small velocity weak transition amplitude $M$ for an elementary
particle $A$. Let it be the scalar vertex, that is, mediated by the
scalar current $J$, $M\!=\!\matel{A(v')}{J(0)}{A(v)}$. We consider it in a
frame moving with small velocity $\vec{v}$. To simplify consideration,
we can assume that the change in velocity $\Delta \vec {v}$ is even
smaller: $|\vec{v}\,'\!-\!\vec{v}\,| \!=\! |\Delta \vec {v}\,| \!\ll\! 
|\vec {v}\,| \ll 1$. For scalar particle $A$ the transition amplitude is
described by a single formfactor, so that in this kinematics
\beq
\matel{A(\vec{v}\,')}{J(0)}{A(\vec{v}\,)} \simeq \mbox{const}(1-
a\,(\delta\vec{v}\vec{v}\,)\,)
\;.
\label{qu5}
\eeq
If acceleration proceeds in the direction transverse to the velocity, 
this amplitude does not depend on the absolute velocity of the
particle.

For spin-$\frac{1}{2}$ particle the scalar amplitude likewise is
described by a single formfactor, however the nonrelativistic amplitudes 
has the new structure:
\beq
\matel{A(\vec{v}\,')}{J(0)}{A(\vec{v}\,)} \simeq
\mbox{const}\,(\varphi^\dagger \varphi - 
\mbox{$\frac{1}{4}$}\, i[\delta \vec{v} \mbox{{\scriptsize $\times$}} 
\vec{v}\,]\!\cdot\!\varphi^\dagger 
\vec\sigma\varphi
- a\,(\delta\vec{v}\vec{v}\,)\,)
\label{qu7}
\eeq
which depends on the velocity of the particle. The similar structure
antisymmetric in $\vec{v}$ and $\delta\vec{v}$ is present for any
particle with nonzero angular momentum. This amplitude is remarkable
since does not depend on the internal structure of the particle: it
can be an elementary pointlike object, or a bound state. It depends 
only on its spin. Actual dynamics
affects only the part of the amplitude symmetric in $\vec{v}$ and
$\Delta\vec{v}$. 
The origin of the antisymmetric term is simply the transformation 
properties of
the spin wavefunctions and roots to the noncommutativity of Lorentz
boosts $U(\vec{v}\,)$ applied in different directions.

Let us illustrate this point. The state $\state{A(\vec{v}\,)}$ can be
obtained boosting the rest-frame particle, $U(\vec{v}\,)
\state{A(0)}$; likewise $\state{A(\vec{v}\!+\!\Delta \vec{v})} =
U(\vec{v}\!+\!\Delta \vec{v}\,) \state{A(0)}$. Nontrivial dependence of
the overlap $\langle A(\vec{v}\!+\!\Delta \vec{v}\,) \state{A(\vec{v}\,)}$
on the frame velocity $\vec{v}$ is nothing but the fact that
$U(\vec{v}\!+\!\Delta \vec{v}\,) \ne
U(\Delta\vec{v})\!\cdot\!U(\vec{v}\,)$ in terms bilinear in both
velocities. The commutator of the two infinitesimal boost operators is
the rotation in the $(\Delta \vec{v}, \vec{v}\,)$ plane; its action on a
state amounts to the operator of angular momentum. Thus, the
antisymmetric in $\vec{v}$ and $\Delta\vec{v}$ piece of the transition
amplitude for a particle directly measures its spin.\footnote{Should we
consider a vector current instead of the scalar one, there would be an
additional term related to the Lorentz transformation of the current
itself.} As is clear from the preceding discussion, the piece of
the amplitude we are interested in comes from the phenomenon known as 
{\it Thomas precession}.

Using this general property of the transition amplitudes, we can imagine
measuring separately the heavy quark spin and the total spin of $B^{(*)}$
meson in the following {\it gedanken} experiment which would bring us close to
the sum rule. We start with the $B^{(*)}$ meson at rest and consider
double interaction (scattering) of the external weak current on the $b$
quark, Fig.~6. In the first act the hadron is accelerated to velocity
$\vec{v}$, the second interaction additionally changes velocity by
$\Delta \vec{v}$. Time $t_2\!-\!t_1$ between the two interactions is at our
disposal, and we can vary it. The energy variable $\omega$ conjugated to
$t_2\!-\!t_1$ actually measures the energy of the produced hadrons. For our
purpose we project the final state onto $B^{(*)}$ and, again, select only
the part of the amplitude antisymmetric in $\vec{v}$ and
$\Delta\vec{v}$. 

\begin{figure}[hhh]
 \begin{center}\vspace*{-3mm}
 \mbox{\epsfig{file=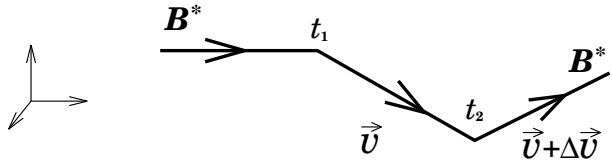,width=8.cm}} \vspace*{-5mm}
  \end{center}
 \caption{
Double scattering process on heavy quark which measures total spin of 
$B^*$ (at large $t_2\!-\!t_1$) or 
spin of $b$ quark (at  $t_2\!\to \!t_1$) via Thomas precession.
}\vspace*{-2mm}
\end{figure}

If we consider the elastic transition where the first act of interaction
produces only moving $B^{(*)}$, the amplitude is given by the total spin
of $B^{(*)}$. This changes if we add other intermediate states. If we sum
the amplitude over {\it all} $b$ hadrons appearing in the intermediate
state, we actually measure only the spin of the heavy quark. Indeed,
this totally inclusive amplitude corresponds to zero time separation
$t_1\!=\!t_2$, so that the light cloud surrounding heavy quark is decoupled
having no time to follow up the scattering. These facts are explicit in
the computation of both the elastic contribution to the scattering
amplitude, and of its OPE expansion \cite{ioffe,newsr}. Therefore, the
contribution of the inelastic states alone directly measures {\it
spin of light degrees of freedom}.

Directing the interested reader to the above original papers for the formal
OPE derivation, here we sketch the quantum mechanical way to obtain
the first sum rule. Our starting point is the expression for the SV
transition amplitudes
\beq
\matel{k(\vec{v}\,)}{J_0(0)}{B^{(*)}(0)} =- 
v_j \frac{\matel{k}{\pi_j}{B^{(*)}}}{\epsilon_k}
\;,
\label{qu9}
\eeq
and similarly for axial currents, up to spin-related factors. Here
$\pi_j\!=\!iD_j$ is the momentum operator of the heavy quark. Below we also 
use the nonrelativistic energy $\pi_0\!=\!iD_0\!-\!m_Q$. The r.h.s.\ in
the above
relation is written in usual quantum-mechanical notations. In the
second-quantized form
\beq
\matel{k(\vec{v}\,)}{J_0(0)}{B^{(*)}(0)} = -v_j
\frac{\matel{k(\vec{v}\!=\!0)}{\bar Q iD_j Q(0)}{B^{(*)}(0)}}{\epsilon_k}
\;.
\label{qu11}
\eeq

Relations for the SV transition amplitudes to the $P$-wave states are most 
easily obtained recalling that they are 
overlaps $\langle k(\vec{v}\,)|B(\vec{v}\!=\!0)\rangle $.
Then we use the general rule
\beq
\state{H_Q(\vec v\,)}\;=\; \state{H_Q(0)} +
\pi_0^{-1}\,(\vec{v}\vec{\pi}\,)\,\state{H_Q(0)}\;+\;{\cal O}(\vec v^{\,2})\;.
\label{v7}
\eeq
which nicely elucidates the meaning of the small velocity sum rules:
the operator $\pi_0^{-1}(\vec v\vec\pi)$ acting on $\state{H_Q}$ is
the generator of the boost along the direction of $\vec v\,$.
Indeed, to get $\state{H_Q(\vec v\,)}$ one must find the eigenstate of
the
Hamiltonian with heavy quark moving with the momentum
$\vec{q}\!=\!m_Q\vec{v}$. The only part which explicitly depends on
momentum comes from  the heavy quark
Hamiltonian $\frac{\vec{\pi}^{2}}{2m_Q}$ (plus
 higher terms in $1/m_Q$). We use the relation
$\exp{(-i\vec{q}\vec{x}\,)}\,{\cal H}_Q\, \exp{(i\vec{q}\vec{x}\,)}=
{\cal H}_Q \!+\!
\vec{v}\vec{\pi} \!+\! m_Q\vec{v}^{\,2}/2$ and drop the last term which
is a
constant; $A_0$ obviously commutes with $x$. Then Eq.\,(\ref{v7})
represents the first-order perturbation theory
in $\delta {\cal H}\!=\! \vec
v\vec \pi$, where $-\pi_0$ plays the role of the unperturbed Hamiltonian 
${\cal H}_0$
(further  details can be  found in Ref.~4 , Eq.\,(178) and
Sect.~VI).
In the second-quantized notations the same relation  takes the
form
\beq
\state{H_Q(\vec v\,)}\;=\; \state{H_Q(0)}\; +\;
\int\; {\rm d}^3\vec{x}\;
\bar Q\,\pi_0^{-1}\,(\vec{v}\vec{\pi}\,)\,Q(x)\state{H_Q(0)}\;+\;
{\cal O}(\vec v^{\,2})\;.
\label{v8}
\eeq

The first sum rule (\ref{ur}), therefore requires computing the
antisymmetric in $j,k$ structure in the following sum
\beq
\sum_n\: \frac{\matel{B^*}{\pi_j}{n}
\,\matel{n}{\pi_k}{B^*}}{(E_n\!-\!E_{B^*})^2}
\;.
\label{qu13}
\eeq
The energies $E_n, E_{B^*}$ are eigenvalues of the total Hamiltonian
\beq
\!\!{\cal H}\;=\;{\cal H}_{\rm light} + {\cal H}_Q 
\mbox{\hspace*{107mm}}
\label{qu15}
\eeq
\vspace*{-7.0mm} 
$$
= {\cal H}_{\rm light}\,-A_0+\frac{1}{2m_Q}\,({\vec\pi}^2 \!+\! \vec\sigma
\vec
B) + \frac{1}{8m_Q^2}\left[-(\vec D\vec E)+ \vec\sigma \!\cdot\!
\{\vec E\!\times\!\vec\pi\!-\!\vec\pi\!\times\!\vec E\} \right] +{\cal
O}(1/m_Q^3),
$$
\vspace*{-2.8mm}\\
where the $1/m_Q^2$ piece is the sum of the Darwin and convection
current ($LS$) terms. In order to compute the sum we, following usual
nonrelativistic quantum mechanics, use the commutation relation
expressing $\vec \pi$ as the commutator of the heavy quark coordinate 
$\vec x$
with the Hamiltonian, but include the $1/m_Q^2$ terms:
\beq
\pi_j= - i\, m_Q \,[x_j,{\cal H}] \:+\:\frac{i}{8m_Q}\left(E_j+
2 \,i \left[\vec\sigma \!\times\!\vec E\right]_j \right)
\;.
\label{qu17}
\eeq
Since each commutator with ${\cal H}$ kills one factor $E_n\!-\!E_{B^*}$
in the denominator, the leading in $m_Q$ term is given by 
$m_Q^2\left(\matel{B^*}{x_j x_k}{B^*} \!-\! 
\matel{B^*}{x_j}{B^*}\,\matel{B^*}{x_k}{B^*}\right)$ which is symmetric
over $j$ and $k$. The antisymmetric part comes in the next order. To
compute it we express, to the leading order in $m_Q$, chromoelectric field 
as the commutator of the heavy quark momentum and the Hamiltonian, 
$E_l\!=\!i[\pi_l,{\cal H}]$. Then we finally get
\bea
\sum_n\: \frac{\matel{B^*}{\pi_j}{n}
\,\matel{n}{\pi_k}{B^*}}{(E_n\!-\!E_{B^*})^2}
= m_Q^2\left[\matel{B^*}{x_j x_k}{B^*} -
\aver{x_j}\,\aver{x_k}\right] 
\nonumber \\ 
& & \mbox{\hspace*{-80mm}} 
+ \mbox{$\frac{i}{8}$} \left[\matel{B^*}{\pi_j x_k \!+\! x_j \pi_k }{B^*}
\!-\!
\aver{\pi_j }\aver{x_k }\!-\!\aver{\pi_k}\aver{x_j}
\right]
\label{qu19} \\
& & \mbox{\hspace*{-80mm}} 
+\mbox{$\frac{1}{4}$}
\left[ 
\matel{B^*}{[\vec\sigma \!\times\! \vec \pi]_j x_k 
\!+\! x_j [\vec\sigma \!\times\! \vec \pi]_k }{B^*}
\!-\!\aver{[\vec\sigma \!\times\! \vec \pi]_j} \aver{x_k} \!-\!
\aver{[\vec\sigma \!\times\! \vec \pi]_k} \aver{x_j}
\right]
\nonumber
\,.
\eea
(The terms with separate expectation values of $\vec{\pi}$ and $\vec{x}$ can 
be discarded.) 
The basic commutator relation $[x_j,\pi_k]=i\delta_{jk}$ thus yields for the
antisymmetric part the matrix element of the heavy quark
spin, $\frac{i}{4} \,\epsilon_{jkl\,}\sigma_{l\,}$. It comes from the 
$LS$ term in the heavy quark Hamiltonian. This
matrix element is given by $\frac{1}{4} \,(\varepsilon'^*_j\varepsilon_k
\!-\!\varepsilon'^*_k\varepsilon_j)$, 
where $\varepsilon$ refer to the polarization vectors of $B^*$.
On the other hand, the explicit summation over the members  of 
the hyperfine $P$-wave multiplets and their polarizations yields for the
antisymmetric part the same
spin structure $(\varepsilon'^*_j\varepsilon_k
\!-\!\varepsilon'^*_k\varepsilon_j)$  
multiplied by $-\tau_{3/2}^2$ and $\tau_{1/2}^2$ for the
$\frac{3}{2}$- and $\frac{1}{2}$-states, respectively 
(see, e.g.\ Ref.~\cite{ioffe}, Sect.4.1). In this way the
general structure of the sum rule (\ref{ur}) is reproduced.

A subtlety must be noted at this point which was previously discussed in the
similar context in Ref.~\cite{optical}. In order to perform the
summation over intermediate states, we had to include in Eq.\,(\ref{qu19}) all
possible intermediate heavy quark states $\state{n}$ which have nonvanishing 
matrix elements. They can be labeled by their explicit excitation number
and the total momentum (not indicated explicitly). As mentioned above, we 
consider the zero-momentum matrix elements, so that the summation over
overall momentum is removed, cf.\ Eq.\,(\ref{v8}). This does not literally
apply, however, to the diagonal transition into the same $B^{(*)}$
states. Although the matrix elements naively contain $\delta$-function
of momentum, the energy denominator in this case amounts to
$(\vec{p}^{\:2}\!/2m_Q)^2$ yielding singularity at small $\vec{p}$. This
singularity must be treated properly and yields a finite contribution.
It must be subtracted from the sum, since the sum rules include only the
transitions into the excited states.

The diagonal contribution can be computed without performing actual 
discretization of the problem, employing the trick used in 
Ref.~\cite{optical}. Considering the infinitesimal momenta of $B^{(*)}$
we can relate the matrix elements of the $b$ quark momentum $\vec\pi$ to the
matrix elements of the total hadron momentum $\vec P\,$:
\beq
\matel{B^{(*)}(\vec{p}\,)}{\pi_j}{B^{(*)}(0)}= \frac{m_Q}{M_{B^{(*)}}}
\matel{B^{(*)}(\vec{p}\,)}{P_j}{B^{(*)}(0)}
\;.
\label{qu21}
\eeq
In the present problem we can neglect the deviation of ratio
$m_Q/M_{B^{(*)}}$ from unity. We then need to evaluate the following sum over
all momenta of $B^{(*)}$:
\beq
\int \frac{{\rm d}^3 \vec{p}}{(2\pi)^3}\,
\frac{\matel{B^*(0)}{P_j}{B^*(\vec{p}\,)}
\matel{B^*(\vec{p}\,)}{P_k}{B^*(0)} } {(E_{B^*}(\vec{p}\,)\!-\!M_{B^*})^2}
\; ;
\label{qu23}
\eeq
This, however, is nothing but the general sum rule (\ref{qu19}) applied
to $B^*$ mesons as (almost) free elementary particles. (The nontrivial 
answer emerges due to {\it ad hoc} interaction assumed to render the spectrum 
discrete.) The antisymmetric
part here amounts then to the expectation value of the spin of $B^*$ as
a whole, and is given by the same polarization structure as above, but
with the additional factor of $2$. Subtracting this contribution from
Eq.\,(\ref{qu19}) finally yields the spin sum rule (\ref{ur}).

The quantum mechanical derivation of the second spin sum rule
(\ref{ur2}) is even simpler if we recall the expressions (\ref{v7}), 
(\ref{v8}) for the moving heavy hadron state. On then has
\beq
\matel{B^*(\vec{v}\,)}{\pi_j}{B^*(0)} \!=\! 
v_k \matel{B^*(0)}{\pi_k \pi_0^{-1} \pi_j}{B^*(0)} \!=\!
-v_k {\sum_n}' \matel{B^*}{\pi_k \pi_0^{-1}}{n} \epsilon_n
\matel{n}{\pi_0^{-1} \pi_j}{B^*}
,
\label{qu29}
\eeq
where all quantum-mechanical states have zero momentum, and the prime 
indicates that summation does not include the diagonal transition. In
writing this we have used the fact that the operator $-\pi_0$ plays the role
of the Hamiltonian for the states with zero momentum. Indeed, due to
the heavy quark equation of motion $\pi_0 Q(x)\!=\!0$ the relation holds for an
arbitrary operator $O$
$$ 
\matel{n}{\bar{Q}\pi_0 O Q(0)}{m} = (E_m\!-\!E_n)\matel{n}{\bar{Q} O Q(0)}{m}
\;.
$$
The transition amplitudes in the sum in Eq.\,(\ref{qu29}) are directly
related to $\tau$'s.  As before, the contributions to the symmetric
and antisymmetric in $j$ and $k$ structures there are
given by $2\tau_{3/2}^2\!+\!\tau_{1/2}^2$ and $\tau_{3/2}^2\!-\!\tau_{1/2}^2$, 
respectively, which yields  Eq.\,(\ref{newsr7}) with $\La$ and 
$\overline\Sigma$ given by sum rules (\ref{vol}), (\ref{ur2}).

The spin sum rule (\ref{ur}) provides the rationale for the experimental fact
that vector mesons $B^*$, $D^*$ are heavier than their
hyperfine pseudoscalar partners $B$, $D$. Indeed, if the sum rule for
$\mu_G^2$ is dominated by
the low-lying states then $\mu_G^2$ must be of the same sign as the
constant in Eq.\,(\ref{ur}), which dictates the negative energy of the
heavy quark spin interaction in $B$ and positive in $B^*$.

The sum rules (\ref{bj})--(\ref{fourthls}) obviously entail a set of exact
QCD inequalities. They are similar to those which have been with us
since the early 1980's\,\cite{EIQCD} and  reflect the most general
features of QCD (such as the vector-like nature of the quark-gluon
interaction). The advent of the heavy quark theory paved the way to a
totally new class of inequalities among the fundamental parameters. As
with the old ones, they are based on the equations of motion of QCD and
certain positivity properties. All technical details of the derivation
are different, however, as well as the sphere of applications.
The first in the series
is the Bjorken inequality $\varrho^2 \!\ge \!\frac{1}{4}$ \cite{BJSR}. 
We, in fact, have a stronger {\it dynamical\,} bound 
$\varrho^2 \!\ge\! \frac{3}{4}$ following from the sum rule (\ref{ur}).
It ensures that at least some of 
the inelastic amplitudes must be nonzero; therefore, the bound state 
with nonzero spin of light cloud cannot be structureless (pointlike) 
regardless of bound-state dynamics. The Bjorken bound is not dynamical
in this respect: the Isgur-Wise function 
$\xi(vv')\!=\!\sqrt{2/(1\!+\!vv')}\,$ with
$\varrho^2 \!=\!\frac{1}{4}$ is the one for the structureless particle.
Comparison to the QCD sum rule evaluation $\varrho^2 \!=\! 0.7\!\pm\!
0.1$ \cite{bloks} suggests that the new bound can be nearly saturated.
We also have the bound $\La\!>\! 2\overline\Sigma$. Other bounds
include $\mu_\pi^2 \ge \mu_G^2$ and $\rho_D^3 \ge |\rho_{LS}^3|/2\,$,
$\,\rho_D^3 \ge -\rho_{LS\,}^3$. We also have direct inequalities between
the parameters of different dimensions, 
\beq 
\mu_\pi^2 \ge \frac{\;3
\La^2}{4\varrho^2 \!-\!1} \;,\qquad \rho_D^3 \ge \frac{3}{8}
\frac{\La^3}{\left(\varrho^2 \!-\!1/4\right)^2} \;,\qquad \rho_D^3 \ge
\frac{\;\;\;\;(\mu_\pi^2)^{3/2}}{\sqrt{3 (\varrho^2 \!-\!1/4)}} \;.
\label{pplrho}
\eeq 
All these inequalities are saturated provided that only one
excited state contributes.

The old nonrelativistic quark models often do not respect the general
relations discussed above. They typically yield small difference
between the $\frac{1}{2}$- and $\frac{3}{2}$- $P$-wave states, and 
therefore hardly can be viewed reliable. A class
of relativistic models was developed recently \cite{leya2} which
incorporate proper Lorentz boost transformations and obey the exact
commutation relations. Consequently, they naturally predict
suppression of the $\tau_{1/2}$ amplitudes compared to
$\tau_{3/2}$. Likewise, they obey the bound $\varrho^2 \! >\!
\frac{3}{4}$. It seems appealing to reexamine the quark
model predictions for processes with heavy flavors in the framework of the
new models \cite{leya2}.

\subsection{Hard QCD and normalization point dependence}

The sum rules (\ref{bj})--(\ref{fourthls}) express 
the heavy quark parameters,
including $\La\!=\! M_B\!-\!m_b$, $\,\mu_\pi^2$
and $\mu_G^2$, as the sum of observable quantities, products of the
hadron mass differences and the transition probabilities. The observable
quantities are scale independent. How then, say, 
$\La\!=\! M_B\!-\!m_b$, $\,\mu_\pi^2$ and $\mu_G^2$ 
happen to be $\mu$-dependent?

The answer is that in  quantum field theory such as QCD the sums
over excited states are generally ultraviolet 
divergent due to physics at  $E \!\gg\! \Lam$. In contrast to 
ordinary quantum mechanics they are not saturated 
by a few lowest states with contributions
fading away fast in magnitude with the excitation number. The 
contributions of hadronic states
with $\epsilon_k \!\gg\! \Lam$ are dual to what we calculate in
perturbation theory using its basic objects, quarks and gluons. The
latter yield the continuous spectrum 
and can be evaluated perturbatively using isolated quasifree heavy
quarks as the initial state. The final states are heavy quarks plus a
certain number of gluons and light quarks. It is the difference between the
actual hadronic and quark-gluon transitions that resides at low excitation
energies.

Therefore, in order to make the sum rules meaningful, we must cut off
the sums at some energy $\mu$ which then makes the expectation values
$\mu$-dependent. The simplest way is merely to extend the sum only up
to $\epsilon_k<\mu$. This is the convention we normally use. 
Thus, in actuality all the sums in the relations (\ref{bj})--(\ref{fourthls})
must include the condition $\epsilon_k < \mu$, which we omitted there
for the sake of simplicity, and all the heavy quark parameters are 
normalized at the scale $\mu$. The exception is the superconvergent
spin-nonsinglet sum rules (\ref{ur}) and (\ref{ur2}) where in the
perturbative domain $\mu \gg \Lam$ such $\mu$-dependence is power
suppressed by factors $\alpha_s\Lam/\mu$ and can be neglected.

\begin{figure}[hhh]
 \begin{center}\vspace*{-2mm}
 \mbox{\epsfig{file=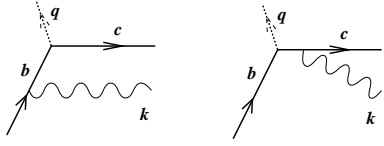,width=5cm}}\vspace*{-5mm}
 \end{center}
\caption{
Perturbative diagrams
determining the high-energy asymptotics of the heavy quark transition
amplitudes and renormalization of the local operators.}
\vspace*{-2mm}
\end{figure}

The high-energy tail of the transitions, to the order $\as$, is given by the
quark diagrams in Figs.~7 with
$$
2\sum_m ... + \sum_n ... \; \ra \; \int \frac{{\rm d}^3\vec k}{2\omega}
$$
where $(\omega,\vec{k})$ is the momentum of the real gluon. The
spin-singlet 
amplitudes are just a constant proportional to $g_s$. Performing 
simple calculations we arrive at the first-order term in the evolution
of, say,  $\mu_\pi^2(\mu)$ \cite{optical},
\beq
\frac{{\rm d}\mu_\pi^2(\mu)}{{\rm d}\mu^2}\; = \; \frac{4}{3}\,
\frac{\as}{\pi} + \ldots
\;.
\label{160}
\eeq
Purely perturbatively, the
continuum analogs of $\tau_{1/2}$ and $\tau_{3/2}$ are equal, and a
similar additive renormalization of $\mu_G^2$ and
$\rho_{LS}^3$ is absent.

The perturbatively obtained evolution equation (\ref{160}) and the 
similar one for the chromomagnetic operator stating its anomalous 
dimension $-3\alpha_s/2\pi$
allow one to determine asymptotics of
$\tau_{1/2}$ and $\tau_{3/2}$  at $\epsilon \gg \Lam\,$,
\bea
2\!\sum_{m} \!\epsilon_m^2|\tau_{3/2}^{(m)}|^2 \!+\!
\sum_{n} \!\epsilon_n^2 |\tau_{1/2}^{(n)}|^2 & \to & 
\;\;\,\frac{8\as(\epsilon)}{9\pi}\: \epsilon\,  {\rm d}\epsilon 
\;,
\label{170}
\\
\;\:\sum_{m}\!
\epsilon_m^2|\tau_{3/2}^{(m)}|^2 \!-\!
\sum_{n}\!
\epsilon_n^2|\tau_{1/2}^{(n)}|^2  &\to & 
-\frac{3\as(\epsilon)}{2\pi} \frac{{\rm d}\epsilon}{\epsilon}
\left[
\sum_{\,\epsilon_m \!<\epsilon\!\!\!} \!\epsilon_m^2|\tau_{3/2}^{(m)}|^2 \!-\!
\sum_{\epsilon_n \!< \epsilon\!\!} \!\epsilon_n^2|\tau_{\!1/2}^{(n)}|^2
\right]\!.
\label{171}
\eea
The last bracket is  $\frac{1}{6}\mu_G^2(\epsilon)$.
Equation (\ref{170}) can be extended to higher orders in $\alpha_s$, this
amounts to using the so-called dipole
coupling $\alpha_s^{(d)}(\epsilon)$ introduced in Ref.~\cite{dipole}:
\beq
\alpha_s^{(d)}(\epsilon) = 
\bar\alpha_s\left({\rm e}^{-5/3+\ln{2}} \epsilon\right)
- 3\mbox{$\,\left(\frac{\,\pi^2\!}{6}\!-\!\frac{13}{12} \right)
\frac{\as^2}{\pi}$} \,+\,{\cal O}(\as^3)\;.
\label{new96}
\eeq
($\bar\alpha_s$ is the standard $\overline{\rm MS}$ strong coupling).
Therefore we get a number of exact perturbative evolution equations
\cite{dipole}
\bea
\mu \frac{{\rm d}\varrho^2(\mu)}{{\rm d}\mu} &=&
\:\frac{8}{9}\:\frac{\alpha_s^{(d)}(\mu)}{\pi}\,, \\
\;\;\frac{{\rm d}\La(\mu)}{{\rm d}\mu} &=&
\frac{16}{9}\,\frac{\alpha_s^{(d)}(\mu)}{\pi} \,,\\
\;\;\frac{{\rm d}\mu_\pi^2(\mu)}{{\rm d}\mu} &=&
\;\frac{8}{3}\:\frac{\alpha_s^{(d)}(\mu)}{\pi}\, \mu\,.
\label{new100}
\eea

Using Eq.\,(\ref{171}) we can estimate the contribution of the high-energy
states in the sum rules (\ref{ur}) and (\ref{ur2}):
\bea
\sum_{\epsilon_m<\mu} \;\;\;|\tau_{3/2}^{(m)}|^2 -
\sum_{\epsilon_n<\mu} \;\;\,|\tau_{1/2}^{(n)}|^2 &\simeq & 
\mbox{\hspace*{.5mm}}\frac{1}{4}\mbox{\hspace*{.5mm}} + 
\frac{\as(\mu)}{8\pi}\: \frac{\mu_G^2(\mu)}{\mu^2}\;,
\label{77}
\\
\sum_{\epsilon_m<\mu} \!\epsilon_m |\tau_{3/2}^{(m)}|^2 -
\sum_{\epsilon_n<\mu} \!\epsilon_n |\tau_{1/2}^{(n)}|^2 &\simeq & 
\frac{\overline\Sigma}{2}+ \frac{\as(\mu)}{4\pi}\:
\frac{\mu_G^2(\mu)}{\mu}
\;;
\label{79}
\eea
they are power suppressed and presumably small in the perturbative
domain.

\subsection{On the saturation of the sum rules}

The question of the saturation of the heavy quark sum rules (in
particular, the lower ones (\ref{bj})--(\ref{pig})) is of primary
importance for phenomenology of the heavy quark expansion. The sum
rules at large enough normalization point $\mu$ tell us what the
asymptotic value on the right-hand side is, and perturbation theory
tells us its $\mu$ dependence. It is a dynamical question starting
from which scale $\mu_0$ this behavior applies. For superconvergent
sum rules (\ref{ur}) and (\ref{ur2}) this is the question at which
scale the sums approach the stated values with a reasonable
accuracy. In order to sensibly apply quantitative $1/m_Q$ expansion,
one must have $m_Q \!\!>\! \!\mu_0$, presumably, $m_Q\!\!\gg\!\!\mu_0$.  While
this is, probably, the case for $b$ particles, such a hierarchy
is not obvious {\it  a priori} in charm.

The existing numerical evaluations of $\La$ and $\mu_\pi^2$ at the
scale near $1\GeV$ suggest rather large values, approximately
$0.7\GeV$ and $0.6\GeV^2$, respectively, which impose rather tight
constraints. These facts were often neglected under various pretexts,
including challenging the accuracy of the QCD sum rules evaluations of
$\mu_\pi^2$. 

Nevertheless, certain constraints following from the 
sum rules are tight and robust simultaneously. 
Namely, the value of $\mu_G^2\!\simeq\!
0.4\GeV^2$, as extracted almost directly
from $B^{(*)}$ and  $D^{(*)}$ masses,
has hardly been challenged. By virtue of the sum rules, the
value of $\mu_\pi^2$ is at least as large. Thus, regardless of the
accuracy in evaluations of the kinetic expectation  value, the
question can be phrased in terms of the generally accepted value of
$\mu_G^2$. At which minimal scale $\mu_0$ the value of
$\mu_G^2(\mu_0)$ reaches $0.3$ or $0.4\GeV^2$? If this scale is below
$1\GeV$, large $\La$ and $\mu_\pi^2$ are almost inevitable. If,
however, $\mu_G^2(1\GeV)$ is significantly below $0.4\GeV^2$, the
chances for success in $1/m_Q$ expansion in charm are slim.

Similar constraints follow from
the sum rule (\ref{ur2}) and, in particular, (\ref{ur}). If these sum
rules are saturated below $1\GeV$, we expect large $\La$ and
$\mu_\pi^2$. If the saturation scale is higher, perturbative treatment
of the scales $\sim \!m_c$ only slightly exceeding $1\GeV$ is not
justified.

The estimates for $\tau_{3/2}$ and $\tau_{1/2}$ for the lowest $P$
wave states group around $0.4\,$, with  
$\epsilon^{(1)}_{3/2}\!\gsim\! \epsilon^{(1)}_{1/2}\!\simeq \!400\:
\mbox{to}\: 500\MeV$ (for the review see Ref.~\cite{fazio}). The old
quark models yielded close values for $\frac{3}{2}$- and 
$\frac{1}{2}$-states.  The more recent relativistic models \cite{leya2} predict
noticeable suppression of the transitions into $\frac{1}{2}$-states
compared to the $\frac{3}{2}$-amplitudes.  Similar absolute values
were reportedly extracted for $\tau_{3/2}$ from the overall 
experimental yield of the
corresponding charmed $P$ wave states \cite{tauexp}. It is evident that
such transitions amplitudes fall short of saturating the sum rules,
$$
\delta^{(1)}_{3/2}\,\mu_G^2 \simeq 0.2\GeV^2\,, \qquad
\delta^{(1)}_{3/2}\, \La \simeq 0.3\GeV\;, \qquad 
\delta^{(1)}_{3/2}\, S_{{\rm light}} \simeq 0.3
$$
($ S_{{\rm light}}$ denotes the sum in Eq.\,(\ref{ur}) for the 
light cloud spin). For the spin-dependent sum rules, the first and the
third entries, the $\frac{1}{2}$-states would further decrease the
values.  In principle, the lowest states alone should not necessarily
saturate the sum rules, even though the idea of the dominance of the
lowest state contributions is very appealing. Let us mention that in
the 't~Hooft model all heavy quark sum rules are saturated with
amazing accuracy by the first excitations \cite{leburkur}. Is such
a possibility excluded in QCD?

Probably, not completely. The dominance of the first excitation with
$\epsilon \!\simeq\! 500\MeV$ (recall that one must use the asymptotic
$m_Q\!\to\infty$ values of the excitation energies and amplitudes) is
still possible if the QCD sum rules underestimate the value of
$\tau^{(1)}_{3/2}$.\footnote{Let us note that the technology of the
QCD sum rules assumes the approximate duality starting
$\epsilon\!=\!1\GeV$ or even lower (the energies are counted from the
heavy quark mass there). Therefore, accepting poor
saturation of the exact heavy quark sum rules at this scale and
simultaneously relying on the QCD sum rules predictions is not
selfconsistent.} Experimental determinations of $\tau$'s are also
questionable since $1/m_c$ corrections are not accounted for
there. The estimates in the 't~Hooft model suggest that they can be
very large. In the cases where they are known explicitly in QCD, the
$1/m_c$ terms generally turn out to be very significant as
well \cite{vain}.

Another -- and, apparently, the most natural -- option is that there
are new states with the masses around $700\MeV$ with similar, or even
larger $\tau^{(2)}_{3/2}\simeq 0.4 \;\mbox{to}\;0.5$; they can be
broad and not identified with clear-cut resonances. The
$\frac{1}{2}$-states must be yet depleted up to this scale. All such
states can be produced in semileptonic $b$ decays and observed as
populating the domain of hadronic invariant mass below or around
$3\GeV$. It will be important to explore these questions in
experiment.

The branching fraction of experimentally identified narrow
$j\!=\!\frac{3}{2}$ $P$-wave states does not exceed a percent level. A
larger fraction was reported recently for the broad hadronic
distributions in $B\to D^{(*)}\pi\,\ell\nu$ decays \cite{exper},
between $2\%$ and $3.5\%$. Such an yield helps to fill the gap
between the total semileptonic fraction $BR_{\rm sl} \simeq 10.5\%$
and identified exclusive channels. Since the observed $D^{(*)}\pi$
mass distribution is very broad, these states were attributed to the
$j\!=\!\frac{1}{2}$ resonances which are expected to have large decay
width. If this is correct, we would rather observe significant
negative contributions to the sum rules (\ref{ur}) and (\ref{pig}) from
the domain of $\epsilon \!\lsim\! 700\MeV$. This possibility does not look
natural from the theoretical viewpoint; neither it is supported by the
predictions of the recent relativistic quark models \cite{leya2,leya}. If it
is nevertheless realized in reality, we would have to expect quite
large values of $\La$ and, in particular, $\mu_\pi^2$.

We note, however, that there is an alternative interpretation of the
data: the reported enhanced yield of broadly distributed $D^{(*)}\pi$
can be the manifestation of nonresonant (continuum) production; then it can
belong to the $\frac{3}{2}$-states. If this is the case, one would
get a consistent picture of conventional saturation of the heavy quark
sum rules at the usual energy scale under $1\GeV$, without necessity
to invoke new paradigms like a too high onset of quark-hadron duality.

The continuum yield is routinely assumed to be small compared to
the resonant contribution. This, however, is more the heredity of
naive quark models than the fact based directly on QCD. The absolute
branching fraction of this yield is rather small, about $20\%$ of the
overall semileptonic rate, and agrees well with the general fact that
nonresonant contributions are $1/N_c$-suppressed. It seems certain
that such a possibility must be carefully explored before alarming
conclusions are drawn.

Interpreting the data mentioned above can be obscured by the
potentially significant $1/m_c$ effects. The spin of light degrees of
freedom itself is a well defined quantum number only as long as the
limit $m_c \!\to\! \infty$ is considered. While deviations from this
academic case are possibly under control for $D$ and $D^*$, the
situation can be much worse for the excited states, in particular, in
the mass range we discuss. Here the energy of the light cloud already
exceeds $m_c$ itself. The new relativistic quark models \cite{leya2}
can gives us a sense of possible significance of such effects.

\section{${\rm BR}_{\rm \,sl}(B)$ and $n_c$}

Before concluding, I would like to mention another nonstandard
possibility which can affect the analysis of ${\rm BR}_{\rm \,sl}(B)$ and the
average number of charm quarks per $B$ decay $n_c$. As explained
elsewhere, the problem of semileptonic fraction is usually considered in
conjunction with $n_c$ -- the latter constrains the total decay rate in
the $b\to c\bar{c}\,s(d)$ channel and thereby allows to isolate potential
uncertainties in the latter due to limited energy release (see, e.g.,
Ref.~\cite{ioffe}, Sect.~7.1).

The point is that there is a certain assumption in the standard analysis
which goes beyond applying the OPE {\it per se}. Namely, the partial
decay rate mediated by quark transition $b\to c\bar{c}s$ and computed
via the OPE, is implicitly equated to the probability of observing
hidden charm, or open charm plus anticharm states in the final state.
Stating differently, the charmless final states and those having a
single charm particle are presumed to be totally independent of the $b\to
c\bar{c}s$ quark level transitions. While this is certainly the case for
the single-charm channels, it is an additional assumption for the
charmless decays; the possibility that it is violated was considered in 
Refs.~\cite{dunietz}. 

Since charm quark is heavy in the scale of strong interactions, 
it is more than plausible 
that this assumption is accurate enough. Yet it must be
realized that we do not derive it from the first principles of QCD. In
particular, the OPE does {\em not} tell us this.\footnote{More
precisely, I am not aware of the way to justify it from the OPE {\it per
se}, without recourse to additional assumptions.} The OPE predicts the
total decay probability induced by the particular term in the weak
Lagrangian, $\bar{c} \gamma_\mu(1\!-\!\gamma_5)b\:
\bar{s}\gamma_\mu(1\!-\!\gamma_5)c$, but it cannot say if it is exhausted
by the final states with two $c$ quarks, or it is shared with the
charmless final states. The OPE width must include all rescattering
processes possible in the final state. Moreover, it is known that
failure to include such contributions can even lead to violation of
general theorem like absence of $\Lam/m_Q$ corrections to the decay
widths \cite{buvmirage}. It is worth emphasizing that we do not
mean the usual effects from the standard Penguin operators -- these
properly describe the corrections to the {\em total} decay rates. Our
point here is the effects of final states interactions which reshuffle
the decay rate between different ``apparent quark channels'', 
but do not change the overall rate.

Since charm quark is relatively heavy, the rescattering probability
between $c\bar{c}$ and light $q\bar{q}$ (or gluonic) states must be
suppressed. Yet even the relatively small fraction can be significant
here. We can introduce the phenomenological parameter ${\cal C}$ to
measure such a probability:
\beq
{\cal C} = \frac{\Gamma_{b\to c\bar{c}s}(B\to\mbox{charmless})}
{\Gamma_{b\to c\bar{c}s}^{\rm tot}}\;,
\label{ch11}
\eeq
where both widths are understood as generated by the the
$(\bar{c}b)(\bar{s}{c})$ or $(\bar{c}b)(\bar{d}{c})$ Lagrangians. (We
assume they are normalized at the scale around $m_b$ to isolate
conventional Penguin operators). By construction, ${\cal C}\!\ne\! 0$ does
not change the total nonleptonic width, but it does affect $n_c$:
\beq
n_c \to n_c- 2\,{\cal C} \!\cdot\! {\rm BR}^{\rm OPE}(B\to c\bar{c}\, X)\;.
\label{ch13}
\eeq
Strictly speaking, ${\cal C}\!\ne\! 0$ would mandate existence of the
opposite processes as well, say, double-charm events in the decays
mediated by $b\to u\bar{u}d$ transitions, beyond the contributions of
the Penguin operators. However, the strong suppression of the KM
disfavored channels makes such effects insignificant.

There exists an experimental bound on such rescattering processes: ${\rm
BR}(B\to\, \mbox{no charm}) \lsim 4\%$ \cite{bound} leads to ${\cal
C}\lsim 0.2$. Even being conservative, we would expect ${\cal C}$ 
smaller than that. (The estimated ${\rm BR}^{\rm OPE}(B\!\to\! 
\mbox{no\,\,charm})$ is below $1\%$.) But the message is clear: Any value of 
${\cal P}\!=\!{\rm BR}^{\rm FSI}(B\to \mbox{no charm})$ from such final state
rescattering would decrease $n_c$ by $2\,{\cal P}$. Therefore, even 
${\cal P}$ about $3\%$
would have the dramatic impact on the allowed domain of theoretical
predictions in the $({\rm BR}_{\rm \,sl}, n_c)$ space. It
therefore seems important to improve the direct experimental bound on
${\rm  BR}(B\to\, \mbox{no\,\,charm})$ down to $1\,\mbox{to}\,2\%$.  
\vspace*{.35cm}\\
{\bf Acknowledgments:}~\,It is a pleasure to thank the
organizers of the Durham Workshop and, in particular, Mike Whalley and Adrian 
Signer for 
hospitality and enjoyable atmosphere. I benefited from discussions with many 
participants, both theorist and experimentalists. Collaboration with
I.~Bigi, M.~Eides, M.~Shifman and A.~Vainshtein is gratefully
acknowledged. I am pleased to thank the authors of Ref.~\cite{leya}, 
especially A.~Le~Yaouanc and L.~Oliver for interesting exchanges
regarding spin sum rules and quark models. 
This work was supported in part by NSF under grant number PHY96-05080
and by RFFI under grant No.\ 99-02-18355.

\end{document}